\documentclass{article}
\usepackage{planargraph}
\usepackage{theoremsetc}
\usepackage{graphicx}
\usepackage{amsmath,amssymb}
\usepackage{algorithm}
\usepackage[noend]{algorithmic}

\newenvironment{proof}
{\paragraph{Proof:}}
{QED\\}

\newcommand{\inflow}{\textrm{inflow}}

\begin{document}

\title{Multiple-Source Multiple-Sink Maximum Flow in Directed Planar Graphs in $O(n^{1.5} \log n)$ Time}

\author{Shay Mozes\\Brown University }

\maketitle

\begin{abstract}
  We give an $O(n^{1.5} \log n)$ algorithm that, given a directed
  planar graph with arc capacities, a set of source nodes and a set of
  sink nodes, finds a maximum flow from the sources to the sinks.  
\end{abstract}

\section{Introduction}
In this paper we give an $O(n^{1.5}\log n)$-time algorithm for the maximum flow
problem in directed planar graphs with multiple sources and
sinks (MSMS). 
\begin{itemize}
\item {\em Input:}
\begin{itemize}
\item a directed planar embedded graph $G$ with non-negative arc capacities
\item a set $S$ of source nodes
\item a set $T$ of sink nodes
\end{itemize}
\item {\em Output:} a feasible flow in  $G$ that maximizes the total flow into $T$.
\end{itemize}

Several new results were recently obtained for this and related problems. A few
months ago Borradaile and Wulff-Nilsen~\cite{BWN10} and Klein and
Mozes~\cite{KM10a} independently presented two $O(n^{1.5} \log n)$-time algorithms for the
maximum flow problem in directed planar graphs with multiple sources
and a single sink (MSSS). Very recently Nussbaum~\cite{Nussbaum10} presented a recursive $O(n^{1.5}
\log^2 n)$-time algorithm for MSMS. His algorithm uses, among other
techniques, MSSS max flow computations.

The proposed algorithm is similar to that of Nussbaum, but uses a simpler
recursive approach. It uses a technique to redistribute excess flow
among a certain set of nodes called boundary nodes. This technique was
used by Nussbaum for the case where the sources and sinks are
separated by the boundary nodes. Here we prove that it 
is applicable in the general case.

\section{preliminaries}
\subsection{Jordan Separators for Embedded Planar Graphs}
Miller ~\cite{Miller86} gave a linear-time algorithm that, given a
triangulated two-connected $n$-node planar embedded graph, finds
a simple cycle separator consisting of at most $2\sqrt{2}\sqrt{n}$ nodes, such
that at most $2n/3$ nodes are strictly enclosed by the cycle, and at
most $2n/3$ nodes are not enclosed.

For an $n$-node planar embedded graph $G$ that is not necessarily triangulated or
two-connected, we define a {\em Jordan separator} to be a Jordan curve $C$
that intersects the embedding of the graph only at nodes
such that at most $2n/3$ nodes are strictly enclosed by the curve and
at most $2n/3$ nodes are not enclosed.  The nodes intersected by the
curve are called {\em boundary nodes}.
To find a Jordan separator
with at most $2\sqrt{2}\sqrt{n}$ boundary nodes, add artificial edges 
with zero capacity to
triangulate the graph and make it two-connected without changing the
maximum flow in the graph. Now apply Miller's
algorithm.

A cycle separator $C$ separates the graph $G$ into two subgraphs
$G_1,G_2$ called {\em pieces}. $G_1$ is the embedded subgraph
consisting of the nodes and edges enclosed by $C$,
i.e. including the nodes intersected by $C$. Similarly, $G_2$ is the subgraph
consisting of the nodes and edges not strictly enclosed by $C$,
i.e. again including the nodes intersected by $C$.

\subsection{Flow}
Let $G$ be a directed graph with arc set $A$, node set $V$, source set
$S \subset V$ and sink set $T \subseteq V \setminus S$.
For notational simplicity, we assume here and henceforth that $G$ has
no parallel arcs and no self-loops.

We associate with each arc $a$ two darts $d$ and $d'$, 
one in the direction of $a$ and the other in the opposite direction.
We say that those two darts are reverses of each other, and write $d =
\rev(d')$.
Given an arc $a$ with capacity $c$, the capacity
associated with the drat $d$ going in the direction of $a$ is
$c$. The capacity associated $\rev(d)$ is zero.

A {\em flow assignment} $f(\cdot)$ is a real-valued function
on darts that satisfies {\em antisymmetry}:
\begin{equation*} \label{eq:antisymmetry}
f(\rev(d))= -f(d)
\end{equation*} 
A flow assignment $f(\cdot)$ is said to {\em respect capacities}
if, for every dart $d$, $f(d) \leq c(d)$. Such a flow assignment is
also called a {\em pseudoflow}.
 
For a given flow assignment $f(\cdot)$, the {\em net inflow} (or just
{\em inflow}) of a node $v$ is $\inflow_f(v) = \sum_{a \in A : \head(a) = v}
f(a) - \sum_{a \in A : \tail(a) = v} f(a)$.
 
A flow assignment $f(\cdot)$ is said to \emph{obey conservation} if
for every node $v \notin S \cup T$, $\inflow_f(v) = 0$.
A pseudoflow that obeys conservation is called a \emph{feasible flow}.
The {\em value} of a feasible flow $f(\cdot)$ is the
 sum of inflows at the sinks, $\sum_{t \in T} \inflow_f(t)$. A {\em
   maximum flow} is a feasible flow whose value is maximum.

The \emph{residual graph} of $G$ with respect to a flow assignment $f(\cdot)$ is the graph $G_f$ with
the same arc-set, node-set, sources and sinks, and with capacity assignment $c_f(\cdot)$
defined by $c_f(d) = c(d) - f(d)$ for every dart $d$.

\section{The Algorithm}
We present the algorithm as a recursive procedure with calls to the following subroutines:
\begin{itemize}
\item  MultipleSourceSingleSinkMaxFlow($G,S,t$) -- computes a maximum flow in $G$ from source set $S$ to sink node $t$. This can be
  implemented in $O(n^{1.5}\log n)$ time using the algorithm
  in~\cite{KM10a} or~\cite{BWN10}.
\item  SingleSourceMultipleSinkMaxFlow($G,s,T$) -- 
 computes a maximum flow in $G$ from source node $s$ to sink set $T$. This can be
  implemented using the multiple-source single-sink maximum flow subroutine.
\item  SingleSourceSingleSinkLimitedMaxFlow($G,s,t,\Delta$) -- computes
  a flow in $G$ from source node $s$ to sink node $t$, whose value is  
  the minimum between $\Delta$ and the value of the maximum $s$-to-$t$ flow. 
 This can be implemented in linear time when $s$ and $t$ are 
  incident to the same face~\cite{Hassin81, HKRS97}.
\end{itemize}

We omit discussion of the
base case of the recursion (the case where the graph size is smaller
than a certain constant).  Each of the recursive
calls operates on a subgraph of the original input graph.  We assume
one global flow assignment $f(\cdot)$ for the original input graph,
and one global capacity assignment $c(\cdot)$. Whenever a subroutine is called, 
it takes as part of its input the  current residual capacity function $c_f(\cdot)$, 
computes a flow assignment  $\widehat f(\cdot)$, and then updates the global flow
assignment $f(\cdot)$ by $f(d) := f(d) + \widehat f(d)$ for every dart
in the subgraph.
In the pseudocode, we do not explicitly mention $f(\cdot)$, $c(\cdot)$
$\sigma(\cdot)$, $c_f(\cdot)$, or $\sigma_f(\cdot)$.
The pseudocode for the algorithm is given below. 

\begin{algorithm}\caption{MultipleSourceMultipleSinkMaxFlow(graph $G$, sources $S$, 
sinks $T$) \label{alg:msms}}
\begin{algorithmic}[1]
\STATE find a simple cycle separator $C$ in $G$ with pieces $G_1$ and $G_2$
\FOR {$i=1,2$}
 \STATE MultipleSourceMultipleSinkMaxFlow($G_i$, $S\cap G_i$, $T\cap G_i$) \label{line:recurse}
 \STATE add to $G_i$ artificial bi-directional arcs with infinite
 capacity between the boundary nodes and an artificial node $v^\star$
 embedded in the face of $G_i$ resulting from the deletion of arcs of
 $G$ not in $G_i$. 
 \STATE MultipleSourceSingleSinkMaxFlow($G_i$,$S\cap G_i$,$v^\star$) \label{line:sources-to-C}
 \STATE SingleSourceMultipleSinkMaxFlow($G_i$,$v^\star$,$T\cap G_i$) \label{line:C-to-sinks}
 \STATE remove $v^\star$ and the artificial arcs from $G_i$ 
\ENDFOR
\STATE Let the nodes of $C$ be $p_1, p_2, \ldots, p_k$
\FOR {$i=1,2, \dots, k$} 
 \STATE Add infinite capacity artificial bi-directional arcs
 $p_{i+1}p_{i+2}$, $p_{i+2}p_{i+3}$, $\ldots$,  $p_{k-1}p_k$
 \IF {$\inflow(p_i) >0$}
  \STATE SingleSourceSingleSinkLimitedMaxFlow($G,p_i,p_{i+1},\inflow(p_i)$) \label{line:fix-plus}
 \ELSE
  \STATE SingleSourceSingleSinkLimitedMaxFlow($G,p_{i+1},p_i,|\inflow(p_i)|$) \label{line:fix-minus}
 \ENDIF
 \STATE remove artificial arcs
\ENDFOR
\STATE push flow from boundary nodes with positive inflow back to
sources and to boundary nodes with negative inflow from sinks\label{line:pseudo}
\end{algorithmic}
\end{algorithm}

The algorithm first finds a Jordan separator $C$ with pieces $G_1$
and $G_2$. For each piece, it calls itself recursively
(Line~\ref{line:recurse}), so that after the call there are no
$S$-to-$T$ residual paths within $G_i$. 
It then pushes flow from the sources in $G_i$ to the boundary nodes using a
multiple-sources single-sink max flow computation (Line~\ref{line:sources-to-C}), and similarly
pushes flow from the boundary nodes to the sinks in $G_i$
(Line~\ref{line:C-to-sinks}).
After the first loop terminates, there are no $S$-to-$C$ residual
paths, no $C$-to-$T$ residual paths and no $S$-to-$T$ residual paths
in the entire graph. The resulting flow assignment is
a pseudoflow rather than a feasible flow since it does not satisfy conservation at the boundary nodes.

The algorithm then handles the boundary nodes one by one in cyclic order.
Along the iterations of the second loop, we say that a boundary node $p_j$ is
\emph{processed} if the current value of the loop variable $i$ is greater
than $j$, and unprocessed otherwise. 
If the inflow at an unprocessed node $p_i$ is positive, the algorithm
tries to resolve the violation of conservation at $p_i$ by sending the excess
flow from $p_i$ to other unprocessed
nodes. Similarly, if the inflow at $p_i$ is negative,  the algorithm
tries to send flow from other unprocessed nodes to $p_i$. This
approach for resolving excess flow in the boundary nodes was very
recently used in~\cite{Nussbaum10} for the special case where all
sources are in one piece and all sinks are in the other. We
prove that the procedure works even in the general case where the
sources and sinks are in both pieces. 

After all boundary nodes are processed, the  algorithm converts the
pseudoflow to a maximum feasible flow by sending any remaining excess flow from the boundary back to
the sources and filling flow deficiencies by sending flow to the boundary from the
sinks. This can be done in linear time by first canceling flow 
cycles using the technique of Kaplan and
Nussbaum~\cite{KaplanNussbaum2009}, and then pushing the flow
 in topological sort order (cf.~\cite{KM10a})

\subsection{Correctness}
We will use the following two lemmas in the proof of correctness:
\begin{lemma}(suffix lemma)\label{lem:suf}
Let $f$ be a flow with source set $X$. 
Let $A,B$ be two disjoint sets of nodes.
If there are no $A$-to-$B$ residual paths and no $X$-to-$B$ residual paths
before $f$ is pushed, then there are no $A$-to-$B$ residual paths
after $f$ is pushed.
\end{lemma}
\begin{proof}
$f$ may be decomposed into a cyclic component (a circulation) and an
acyclic component. 
Note that pushing a circulation does not change the amount of flow crossing any
cut. This implies that if there were no
$A$-to-$B$ residual paths before $f$ was pushed, then there are none
after just the cyclic component of $f$ is pushed. 
It therefore suffices to show the lemma for an acyclic flow $f$.

Suppose, for the sake
of contradiction, that there exists
a residual $a$-to-$b$ path $P$ after $f$ is pushed for some $a\in A$
and $b \in B$.
Let $P'$ be the maximal suffix of $P$ that was residual before the
push. That is, the arc $e$ of $P$ whose head is $\pstart(P')$ was
non-residual before the push, and $f(\rev(e)) >
0$. The fact that $f(\rev(e)) > 0$ implies that before $f$ was pushed
there was a residual path $Q$ from some node $x \in X$ to
$\head(e)$. Therefore, the concatenation of $Q$ and
$P'$ was a residual
$x$-to-$b$ residual path before the push, a contradiction.
\end{proof}

\begin{lemma}(prefix lemma)\label{lem:pref}
Let $f$ be a flow with sink set $X$. 
Let $A,B$ be two disjoint sets of nodes.
If there are no $A$-to-$B$ residual paths and no $A$-to-$X$ residual paths
before $f$ is pushed, then there are no $A$-to-$B$ residual paths
after $f$ is pushed.
\end{lemma}
\begin{proof}
Similarly to the proof of Lemma~\ref{lem:suf}, it suffices to prove the lemma for an
acyclic flow $f$.
Suppose, for the sake
of contradiction, that there exists
a residual $a$-to-$b$ path $P$ after $f$ is pushed for some $a\in A$
and $b \in B$.
Let $P'$ be the maximal prefix of $P$ that was residual before the
push. That is, the arc $e$ of $P$ whose tail is $\pend(P')$ was
non-residual before the push, and $f(\rev(e)) >
0$. The fact that $f(\rev(e)) > 0$ implies that before $f$ was pushed
there was a residual path $Q$ from $\tail(e)$ to some node $x \in X$.
Therefore, the concatenation of $P'$ and $Q$ was a residual
$a$-to-$X$ residual path before the push, a contradiction.
\end{proof}

As Nussbaum points out~\cite{Nussbaum10}, we may assume, without loss of
generality, that no sources or sinks belong to $C$. Otherwise we may
replace each such terminal (i.e., either a source or a sink) $v$ with a new terminal $v'$ embedded in a
face to which $v$ is adjacent, connect $v'$ to $v$ and designate
$v'$ as the terminal instead of $v$. 

\begin{lemma}\label{lem:S-T-Gi}
After Line~\ref{line:recurse} is executed for piece $G_i$, there is
no $S$-to-$T$ residual path in $G_i$.
\end{lemma}
\begin{proof}
By maximality of the flow pushed in Line~\ref{line:recurse}
\end{proof}

\begin{lemma}\label{lem:S-C}
After Line~\ref{line:sources-to-C} is executed for piece $G_i$, there is
no:
\begin{enumerate}
\item $S$-to-$T$ residual path in $G_i$ \label{SCST}
\item $S$-to-$C$ residual path in $G_i$ \label{SCSC}
\end{enumerate}
\end{lemma}
\begin{proof}
Item~\ref{SCSC} follows from the maximality of the flow pushed in Line~\ref{line:sources-to-C}.
Item~\ref{SCST} follows by applying the suffix lemma with $A=S\cap
G_i$, $B=T\cap G_i$, $X=S\cap G_i$, and $f$ the flow pushed in Line~\ref{line:sources-to-C}.
\end{proof}

\begin{lemma}\label{lem:C-T}
After Line~\ref{line:C-to-sinks} is executed for piece $G_i$, there
is no:
\begin{enumerate}
\item $S$-to-$T$ residual path in $G_i$\label{CTST}
\item $S$-to-$C$ residual path in $G_i$\label{CTSC}
\item $C$-to-$T$ residual path in $G_i$\label{CTCT}
\end{enumerate}
\end{lemma}
\begin{proof}
Item~\ref{CTCT} is immediate from the maximality of the flow
pushed in Line~\ref{line:C-to-sinks}.
Item~\ref{CTST} follows by applying the prefix lemma with $A=S\cap
G_i$,
$B=T \cap G_i$, $X=T \cap G_i$, and $f$ the flow pushed in Line~\ref{line:C-to-sinks}.
Item~\ref{CTSC} follows by applying the prefix lemma with $A=S\cap
G_i$,
$B=C$, $X=T \cap G_i$, and $f$ the flow pushed in Line~\ref{line:C-to-sinks}.
\end{proof}

An immediate corollary of Lemma~\ref{lem:C-T} is
\begin{corollary}\label{cor:init}
Immediately after the first loop terminates there is no:
\begin{enumerate}
\item $S$-to-$T$ residual path in $G$
\item $S$-to-$C$ residual path in $G$
\item $C$-to-$T$ residual path in $G$
\end{enumerate}
\end{corollary}

\begin{lemma}\label{lem:invs}
The following invariants are preserved throughout the execution of the
second loop
\begin{enumerate}
\item There is no $S$-to-$T$ residual paths in $G$.\label{inv:ST}
\item There is no residual $S$-to-$C$ path nor residual $C$-to-$T$
  path in $G$.\label{inv:C}
\item If a processed node $p_j$ has positive inflow, there is no
  residual path from $p_j$ to the as-yet-unprocessed nodes. If $p_j$
  has negative inflow, there is no residual path to it from the as-yet-unprocessed nodes.\label{inv:proc1}
\item There is no residual path from a processed node with positive
  inflow to a processed node with negative inflow.\label{inv:proc2}
\end{enumerate}
\end{lemma}
\begin{proof}
By induction on the number of iterations $i$ of the loop (i.e., the
number of processed nodes).
By corollary~\ref{cor:init}, the first two invariants are satisfied
immediately before the second loop is executed. The last two
invariants are trivially satisfied since at that time there are no
processed nodes.

Assume the invariants hold up until the beginning of the $i^{th}$
iteration. Suppose that $p_i$ has positive inflow at the beginning of
the iteration (the case of
negative inflow is similar). At the end of the iteration $p_i$ is
processed and we need to show that the invariants still hold. 
\begin{enumerate}
\item Invariant~\ref{inv:ST} holds by invoking
the suffix lemma with $A=S$, $B=T$, $X=\{p_i\}$, and $f$ the flow
pushed from $p_i$ in Line~\ref{line:fix-plus}. 
\item There are no residual $S$-to-$C$ paths by invoking the prefix lemma
with $A=S$,  $B=C$, $X=\{p_j : j>i\}$ and $f$ the flow
pushed from $p_i$ in Line~\ref{line:fix-plus}. 
There are no residual $C$-to-$T$ paths by invoking the suffix lemma
with $A=C$,  $B=T$, $X=\{p_i\}$ and $f$ the flow
pushed from $p_i$ in Line~\ref{line:fix-plus}.
\item Since $p_i$ had positive inflow at the beginning of the
  iteration, and the flow pushed in Line~\ref{line:fix-plus} is
  limited, if $p_i$ has non-zero inflow at the end of the iteration it
  must be positive, and the flow pushed was in fact a maximum flow
  from $p_i$ to $\{p_j:j>i\}$.
  Invariant~\ref{inv:proc1} holds for $p_i$ by maximality of the flow
pushed  in Line~\ref{line:fix-plus}.
The invariant holds for processed nodes $p_j$ with
  $j<i$ by invoking the prefix lemma with $A=\{p_j:j<i\}$, $B =
  \{p_j:j>i\}$, $X = \{p_j:j>i\}$, and $f$ the flow
pushed from $p_i$ in Line~\ref{line:fix-plus}.
\item Invariant~\ref{inv:proc2} holds for $\{p_j:j<i\}$ by invoking
  the prefix lemma with $A = \{p_j : j<i, \inflow(p_j)>0\}$, $B =
  \{p_j : j<i, \inflow(p_j)<0\}$, $X = \{p_j:j>i\}$, and $f$ the flow
pushed from $p_i$ in Line~\ref{line:fix-plus}. The invariant holds for
$p_i$  by invoking the suffix lemma with $A=\{p_i\}$, $B =
  \{p_j : j<i, \inflow(p_j)<0\}$, $X = \{p_i\}$ and  $f$ the flow
pushed from $p_i$ in Line~\ref{line:fix-plus}.
\end{enumerate}
\end{proof}

\begin{theorem}\label{lem:final}
The flow computed by the algorithm is a maximum feasible flow.
\end{theorem}
\begin{proof}
The flow pushed by the algorithm originates only at sources and
boundary nodes and terminates only at sinks and boundary
nodes. Therefore, sources, sinks and boundary nodes are the only nodes
whose inflow might be non-zero.  
Since line~\ref{line:pseudo} makes the inflow at all boundary nodes
zero, the flow assignment upon termination is a feasible flow. 
It remains to show that upon termination there is no residual $S$-to-$T$ path.
Let $C_+$ ($C_-$) be the set of nodes with positive (negative) inflow
just before Line~\ref{line:pseudo} is executed.
Let $f_+$ ($f_-$) be the flow pushed back from $C_+$ to $S$ (from $T$
to $C_-$) in Line~\ref{line:pseudo}.
Consider first pushing back $f_+$.
By Lemma~\ref{lem:invs}, we may invoke the suffix lemma with $A=S \cup
C_-$, $B=T$,
$X=C_+$ and $f=f_+$ to show there are no $S$-to-$T$
residual paths nor $C_-$-to-$T$ residual paths after $f_+$ is pushed. 
Similarly, invoking the suffix lemma with $A=S$, $B=C_-$,
$X=C_+$ and $f=f_+$ shows there is no $S$-to-$C_-$
residual path after $f_+$ is pushed. 
Next, consider pushing $f_-$.  
Invoking the prefix lemma with $A=S$, $B=T$, $X=C_-$ and $f=f_-$
shows there are no residual $S$-to-$T$ paths after
Line~\ref{line:pseudo} is executed.
\end{proof}

\subsection{Running Time}
The algorithm performs one recursive call per piece. In addition, it
performs two
multiple-source single-sink max flow computations per
piece, which take $O(|G_i|^{1.5}\log n)$ time. 
Processing the boundary nodes takes $O(n)$ per node since the maximum
flow computed is between a source and a sink on the same face.
Therefore, the non-recursive part takes $O(n^{1.5}\log n)$.
Since the size of the two pieces (including boundary nodes) are at most $\theta n + 2\sqrt{2}\sqrt{n}$
and $(1-\theta) n + 2\sqrt{2}\sqrt{n}$, for some $1/3 \leq \theta \leq
2/3$, the total running time is thus bounded by
\begin{equation}\label{eq:rec}
T(n) \leq c_1 n^{1.5} \log n + \max_{1/3 \leq \theta \leq 2/3} \left\{T(\theta n + \sqrt{8n}) +
T((1-\theta) n + \sqrt{8n})\right\}
\end{equation}
for some constant $c_1$.

\begin{lemma} $T(n) = O(n^{1.5} \log n)$ \end{lemma}
\begin{proof} We prove, by induction on $n$, that there exists a constant $c$ such that for sufficiently large $n$,  $T(n) < cn^{1.5} \log n$.
We choose $c$ sufficiently large so that the base of the induction, where $n$ is some constant, holds.
For the inductive step, using the inductive hypothesis in
Eq.~\ref{eq:rec} we get 
$$T(n) \leq  n^{1.5} \log n \left[ c_1 + \max_{1/3 \leq \theta \leq 2/3} \left\{\left(\theta
+ \sqrt {8/n}\right)^{1.5} + \left(1-\theta +   \sqrt
{8/n}\right)^{1.5}\right\} \cdot c \right]$$
By convexity, the maximum is attained at the extreme values. \\
Since $(1/3)^{1.5} + (2/3)^{1.5} = 0.7376\dots$,  for sufficiently large $n$, $\left(\theta
+ \sqrt {8/n}\right)^{1.5} + \left(1-\theta + \sqrt{ 8/n}\right)^{1.5} < 0.74$. Therefore, $c$ can be chosen
sufficiently large so that $c_1 + 0.74c < c$, proving the lemma.
\end{proof}

We note that the running-time bottleneck are the multiple-source single-sink
max flow computations. An $O(n^{1.5})$ bound on MSSS would imply, by the above
proof an $O(n^{1.5})$ bound for multiple-sources multiple-sinks
maximum flow as well.

\section*{Acknowledgements}
I acknowledge fruitful discussions with Cora Borradaile, Philip Klein, Yahav Nussbaum and Christian Wulff-Nilsen.
\bibliographystyle{plain}
\bibliography{long,all}

\begin{thebibliography}{1}

\bibitem{BWN10}
Glencora Borradaile and Christian Wulff-Nilsen.
\newblock Multiple source, single sink maximum flow in a planar graph.
\newblock {\em CoRR}, abs/1008.4966, 2010.

\bibitem{Hassin81}
R.~Hassin.
\newblock Maximum flow in $(s,t)$ planar networks.
\newblock {\em Information Processing Letters}, 13:107, 1981.

\bibitem{HKRS97}
M.~R. Henzinger, P.~N. Klein, S.~Rao, and S.~Subramanian.
\newblock Faster shortest-path algorithms for planar graphs.
\newblock {\em Journal of Computer and System Sciences}, 55(1):3--23, 1997.

\bibitem{KaplanNussbaum2009}
Haim Kaplan and Y.~Nussbaum.
\newblock Maximum flow in directed planar graphs with vertex capacities.
\newblock In {\em ESA 2009}, pages 397--407, 2009.

\bibitem{KM10a}
Philip~N. Klein and Shay Mozes.
\newblock Multiple-source single-sink maximum flow in directed planar graphs in
  ${O}(n^{1.5} \log n)$ time.
\newblock {\em CoRR}, abs/1008.5332, 2010.

\bibitem{Miller86}
G.~L. Miller.
\newblock Finding small simple cycle separators for 2-connected planar graphs.
\newblock {\em Journal of Computer and System Sciences}, 32(3):265--279, 1986.

\bibitem{Nussbaum10}
Yahav Nussbaum.
\newblock Multiple-source multiple-sink maximum flow in planar graphs.
\newblock {\em CoRR}, abs/1012.4767, 2010.

\end{thebibliography}

\end{document}